\documentclass[%
 reprint,
 amsmath,amssymb,
 aps,
]{revtex4-2}

\usepackage{xcolor}

\usepackage{mathtools}
\usepackage{array}
\usepackage{url}

\begin{document}

\title{How Thin and Efficient Can a Metasurface Reflector Be? Universal Bounds on Reflection for Any Direction and Polarization}

\author{Mohamed Ismail Abdelrahman}
 \email{mia37@cornell.edu}
 \affiliation{School of Electrical and Computer Engineering, Cornell University, Ithaca, NY, USA }%
 
\author{Francesco Monticone}
 \email{francesco.monticone@cornell.edu}
  \affiliation{School of Electrical and Computer Engineering, Cornell University, Ithaca, NY, USA }%

\begin{abstract}
Light reflection plays a crucial role in a number of modern technologies. In this paper, analytical expressions for maximal reflected power in any direction and for any polarization are given for generic planar structures made 
of a single material represented by a complex scalar susceptibility. 
The problem of optimal light-matter interactions to maximize reflections is formulated as the solution of an optimization problem in terms of the induced currents, subject to energy conservation and passivity, which admits a global upper bound by using Lagrangian duality. 
The derived upper bounds apply to a broad range of planar structures, including metasurfaces, gratings, homogenized films, photonic crystal slabs, and more generally, any inhomogeneous planar structure irrespective of its geometrical details. These bounds also set the limit on the minimum possible thickness, for a given lossy  material, to achieve a desired reflectance. Moreover, our results allow discovering parameter regions where large improvements in the efficiency of a reflective structure are possible compared to existing designs. Examples are given of the implications of these findings for the design of superior and compact reflective components made of real, imperfect (i.e., lossy) materials, such as ultra-thin and efficient gratings, polarization converters, and light-weight mirrors for solar/laser sails.
\end{abstract}

\maketitle






\section{Introduction}

Advanced nano-fabrication and nano-patterning techniques allow unprecedented control over light-matter interactions, opening vast opportunities for light-based technologies with enhanced performance. Nanophotonic solar cells are a good example of how confining light at subwavelength scales can dramatically increase the absorption enhancement beyond the conventional limit (Yablonovitch limit) for bulky solar cells \cite{yu2010fundamental}. Subwavelength patterning allows to realize not only better optical devices, but also novel effects and functions that were not previously thought to be possible with natural materials, such as negative refraction, invisibility and artificial magnetism, light manipulation over thin surfaces \cite{pendry2000negative,shelby2001experimental,smith2004metamaterials,pendry2006controlling,ginn2012realizing,yu2014flat}. 

Because of the capacity to create nano-structures in virtually unlimited forms and with high precision, the design space of all conceivable geometrical configurations for a given volume is vast, possibly spanning thousands to millions of optimization variables. Design methods that involve large-scale simulations, either via brute-force parametric sweeps or using more advanced inverse-design and optimization algorithms, are the typical approach to arrive at components with superior performance. There is no doubt that this approach is successful in achieving efficient designs \cite{rahmat1999electromagnetic,lu2013nanophotonic,piggott2017fabrication,minkov2020inverse}; however, it suffers from a fundamental weakness: no matter how many simulations are performed, there is typically no guarantee that a globally maximal/minimal solution could be identified. If a structure's performance metrics are already close to the optimal solution, significant computing work could be wasted for the sake of finding a better design with no noticeable enhancement. In this context, the question of how to determine a fundamental bound on a certain physical response in a specific volume (such as absorption, scattering, reflection, etc.), no matter how finely structured the system is, has become critically important both from a scientific and a practical perspective. To obtain universal bounds on optical response, these questions should be approached from a fundamental perspective, by examining basic physics constraints like energy conservation, causality, passivity, and symmetries, which govern the totality of electromagnetic interactions. Several fundamental bounds have already been identified in the literature, such as bounds on the scattering cross-section, absorption, near-field radiative heat transfer, antenna performance, local density of states, refractive index, and other physical responses  \cite{gustafsson2012optimal,gustafsson2020upper,shim2019fundamental,kuang2020maximal,shim2021fundamental,chao2022physical}.  In a design approach informed by fundamental bounds, inverse-design methods can then be used to determine an actual feasible design as close as possible to the global optimum.

A particularly significant optical function that has only received marginal attention in this context is the ability to optimally control the reflected power in terms of magnitude, phase, direction, and polarization, which is important for a plethora of applications, from standard ones in the context of reflection gratings, polarizers, and mirrors, to more exotic scenarios. For instance, maximizing and controlling reflections with the smallest possible amount of material is of crucial importance for solar/laser sails powered by radiation pressure. A lighter solar sail can more easily be accelerated to higher speeds, potentially reaching a significant fraction of the speed of light \cite{vulpetti2014solar,davoyan2021photonic}. Furthermore, channeling the incident light power into the orthogonal polarization through reflection is critical to create chiral cavity modes (modes with well-defined handedness), which in turn can enhance molecular detection sensitivity \cite{kang2017preserving,feis2020helicity}. Many applications, such as holography, light-based radars (lidars), virtual/augmented reality, and solar sail steering, also require the ability to reflect power in a specific direction with very high efficiency using thin platforms (e.g., metasurfaces) \cite{davoyan2021photonic,zheng2015metasurface,li2019phase,lio2021lidar,lesina2020tunable}.

\begin{figure}[h!] 
\centering\includegraphics[width=8cm]{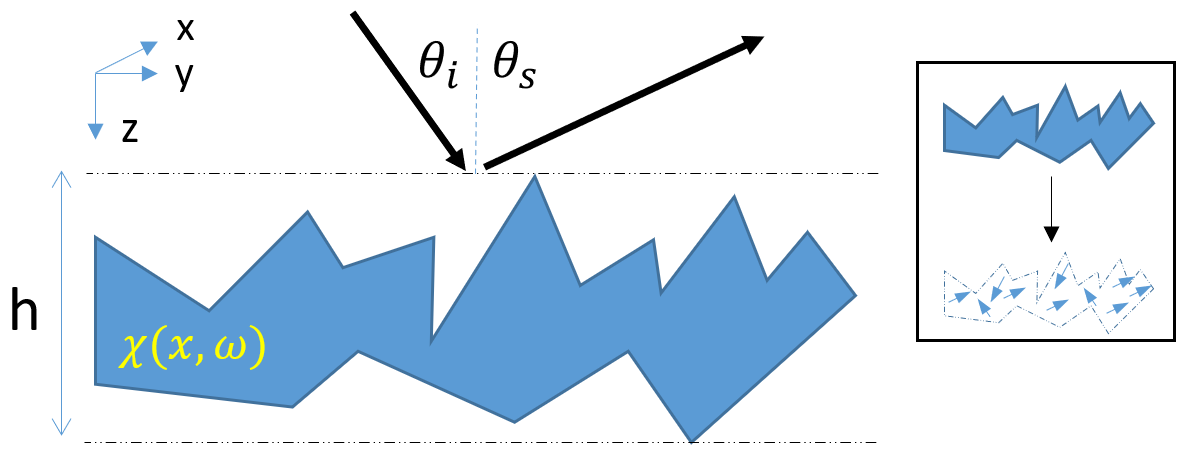}
\caption{Our goal is to find an upper bound on the general reflection problem from a generic planar structure of thickness $h$ (incidence and reflection directions and polarizations are arbitrary, and there is no prior assumption on the optimal patterning details). The bound indicates the highest possible reflection over all possible material distributions for a single material characterized by a complex polarizability $\chi$. For simplicity, the structure width is assumed to be much larger than the thickness $h$, and the considered material is local, isotropic, and nonmagnetic. (Inset): The optimization problem is formulated using induced polarization currents radiating in vacuum, which results in the same fields originating from the inhomogeneous material distribution. To find an analytical expression for the bound, all the possible induced current distributions are assumed to be bounded by a film of thickness $h$ and centered at $z=0$} \label{Fig1}
\end{figure}

In this paper, we determine analytical closed-form expressions for the upper bound on the reflected power in any direction and polarization from a generic planar structure of thickness $h$ that is made of a single material characterized by a complex scalar susceptibility $\chi$. No prior assumption about the structural features and details of the optimal design is assumed, as illustrated in {Figure \ref{Fig1}}. The methodology used to derive the upper bounds is adopted from Ref. \cite{kuang2020maximal}, where the authors developed a comprehensive framework using convex optimization techniques to determine global bounds on single-frequency light-matter interactions constrained by energy conservation, manifested in the so-called ``Poynting's theorem.'' Intuitively, this theorem shows how the combined action of both scattering and absorption processes restricts the induced polarization currents $\bf{\Phi}$, and accordingly, the physical response of the structure. Using this approach, Ref. \cite{kuang2020maximal} derived fundamental upper bounds for the total extinction, scattering, and absorption cross sections of a material body. When compared to earlier bounds that relied solely on either the scattering or absorption processes to construct energy-conservation constraints, the bounds derived in Ref. \cite{kuang2020maximal} have been shown to be tighter and converge at all scales.

Based on this approach, the bounds are derived from an optimization problem formulated in terms of the induced currents, with a general form given as follows,

\begin{flalign} \label{eq:1}
  \begin{aligned}
&\max\,\,\,\,\,\,\,\, \bf{\Phi^{\dagger}}  {\overline{A}} \bf{\Phi} \\
    &\textrm{given}\,\,\,\,\,\, \textrm{Im}(\bf{E}^i \bf{\Phi})=\bf{\Phi^{\dagger}}  (\textrm{Im}\, \zeta + \textrm{Im}\, \bf{\Gamma_0}) \bf{\Phi} 
  \end{aligned}&&&
\end{flalign}

 In this vector notation, the inner product between two vectors  is defined as the integral over the structure volume: $\bf{X^{\dagger}Y}=\int_V \bf{X(r')} \cdot \bf{ Y(r')}\,  dV' $, while matrix products act as convolution operators (the integrals and position dependencies may also be entirely removed by assuming any sufficiently good numerical discretization of the problem, in which case $\bf{X}$ and $\bf{Y}$ would be vectors of size $p N\times 1$ for $N$ spatial degrees of freedom and $p$ polarization degrees of freedom). The objective function to be maximized represents a general quadratic power-flow response function, defined by the matrix $\bf{{\overline{A}}}$ (further details below), while the constraint equation is the volume-integral form of the energy conservation law (Poynting's theorem for real power conservation) that governs the light-matter interaction: the total power extracted from the excitation wave $\bf{E}^i$ is equal to the sum of total absorbed and scattered power by the material, represented respectively by the loss term $\textrm{Im}~ \zeta = \textrm{Im}~ \chi / |\chi|^2$ and the electromagnetic free-space Green's function tensor $\textrm{Im}~ \bf{\Gamma_0}$. Here, and in the rest of the paper, passivity is assumed, i.e., no gain mechanism exists in the medium.

The crucial advantage of formulating the problem as in Eq.  (\ref{eq:1}) -- a quadratic objective with a single quadratic constraint -- is that a global maximal solution, i.e., an upper bound, is guaranteed to exist, and it can be readily determined, due to the convexity of the ``dual'' problem \cite{kuang2020maximal}. Moreover, the upper bound of the solution in a given domain $V'_1$ is always the same as or smaller than the upper bound in a larger domain $V'_2\supset	V'_1$ \cite{boyd2004convex}. Therefore, the bound on a slab/film of thickness $h$ that encloses a generic planar structure, as in Figure \ref{Fig1}, necessarily applies to all possible material distributions within the specified volume. Thus, the upper reflection bound on a slab/film, which can be calculated analytically in many cases, is also the upper bound on any complex planar structure of the same (or smaller) thickness. Nevertheless, it is important to note that since energy conservation applies to the given structure as a whole, the bound is not necessarily {tight, namely, it is not necessarily fulfilled by a particular, physical, excitation wave.} Adding more constraints, such as local energy conservation or the reactive version of Poynting's theorem, can result in a tighter bound, but at the expense of no available closed-form solution, {therefore losing relevant physical insight} \cite{gustafsson2020upper,kuang2020computational}.

\section{Upper bounds on the general reflection problem: Derivation}

The physical quantity that represents reflection must meet two conditions for the application of the optimization approach, Eq. (\ref{eq:1}), discussed in the previous section: it must be defined in terms of the induced polarization currents and it must be quadratic (or linear) with respect to those currents. An ideal candidate is the far-field directional radiation intensity, which represents the scattered power in an arbitrarily  direction $\bf{\hat{k}^s}$ and polarization $\bf{\hat{e}^s}$ and can be written in terms of the induced polarization currents as \cite{orfanidis2002electromagnetic,gustafsson2012optimal}

\begin{equation} \label{eq:U}
U({\bf{\hat{k}^s}},{\bf{\hat{e}^s}}) =\frac{k^2}{32 \pi^2}\, \Big| \int_V {\bf{\hat{e}^s}} \cdot {\bf{ J^*(r)}}\, e^{ik{\bf{\hat{k}^s}\cdot r}}\, dV\Big|^2,
\end{equation}
where  $k$ is the free-space wavenumber, and the vacuum permittivity and permeability are set to unity. In this paper, we consider nonmagnetic materials, where only electric polarization currents $\bf{J}(\bf{r})=i\,\omega \bf{\Phi}(\bf{r})$ are nonzero. Using the vector notation, the radiation intensity can then be expressed as
\begin{equation} \label{eq:Uv}
U= \frac{k^4}{32 \pi^2}\, {\bf{\Phi^{\dagger}F^s}}\,{\bf{F^{s\dagger}\Phi}}, 
\end{equation}
where ${\bf{F^s}}=e^{ik{\bf{\hat{k}^s}\cdot r}}\,{\bf\hat{e}^s}$ is the reflection vector indicating the direction and polarization of the reflected wave. The quantity in Eq. (\ref{eq:Uv}) is in the same form as the general objective function in Eq. (\ref{eq:1}), with $\bf{{\overline{A}}}  \propto {\bf{F^s}}\,{\bf{F^{s\dagger}}}$. Thus, using the convexity of the dual problem, and following the same approach as in Ref. \cite{gustafsson2020upper}, \cite{kuang2020maximal}, it can be shown that the upper bound on the directional radiation intensity can be determined as
\begin{equation} \label{eq:Uopt}
U_{\textrm{opt}}= \frac{k^4}{128 \pi^2}\, (\big|\alpha\big|+\sqrt{\beta \gamma})^2,
\end{equation}
where  $\alpha={\bf{F^{s\dagger}} {\overline{G}} \, E^i}$,$\,\beta={\bf{E^{i\dagger}} {\overline{G}}\, E^i}$, $\gamma={\bf{F^{s\dagger}} {\overline{G}}\, F^s}$, while ${\bf{{\overline{G}}}}$ is a matrix given by $(\textrm{Im}\, \zeta + \textrm{Im}\, {\bf  \Gamma_0})^{-1}$. For isotropic materials, $\zeta$ is a scalar multiple of the identity matrix, and $\bf{{\overline{G}}}$ can be directly expanded in terms of the eigenmodes of the Green's function.

As previously discussed, to find analytical expressions for the upper reflection bound for a planar structure of thickness $h$, instead of evaluating Eq. (\ref{eq:Uopt}) for the specific shape of the structure, we evaluate it for a high-symmetry geometry enclosing the structure, i.e., a planar bounding volume of the same thickness, for which analytical expressions for the eigenvectors and eigenvalues of the Green's function are available. {An important point to mention here is that, while we assume that the structure is laterally finite, so that we can properly define a radiation intensity in ``far-field,'' as in (\ref{eq:U}), we nevertheless expand the Green's function using the basis functions for an infinitely extended planar geometry, which are simple propagating plane waves. Clearly, this assumption is valid only if the area of the structure $A$ is much larger than its thickness and the wavelength, which is the case of interest here. Indeed, the validation examples presented in the next section show that this assumption leads to valid results}. 

The expansion of the imaginary part of the Green's function into propagating plane wave modes is given by \cite{kruger2012trace}
\begin{equation}
    \textrm{Im}\, \bf{\Gamma_0} = \sum_{s,p} \int_{k_{\parallel}<k} {\bf{v}}_{s,p}({\bf{k_{\parallel}}}) \, {\bf{v^{\dagger}}}_{s,p}({\bf{k_{\parallel}}}) \, \frac{d {\bf{k_{\parallel}}}}{(2\pi)^2},
\end{equation}
where the index $s$ represents modes with even($+$)/odd($-$) parity, the index $p$ denotes TE or TM polarizations,  ${\bf{k_{\parallel}}}= k_x{\bf{\hat{e_x}}} + k_y{\bf{\hat{e_y}}}$, and $k^2=k_{\parallel}^2+k_z^2$. 
The expressions for the modes $v_{s,p}$ are given in the literature, e.g., in the Appendix of Ref. \cite{kuang2020maximal}.
These modes form a complete and orthogonal set over the film volume as $    {\bf{v^{\dagger}}}_{s,p}({\bf{k_{\parallel}}}) \, {\bf{v}}_{s',p'}({\bf{k'_{\parallel}}})=A\, \rho_{s,p}(k_{\parallel})\, \delta_{ {\bf{k_{\parallel}},k'_{\parallel}}}\, \delta_{s,s'} \, \delta_{p,p'}$, where $\delta$ is the Kronecker delta, and the eigenvalues $\rho_{s,p}(k_{\parallel})$ are given by: $\rho_{\pm,TE}({{k_{\parallel}}})= k^2 h/ k_z\, (1\pm \textrm{sinc}\, k_z\,h )/4$, and $\rho_{\pm,TM}(k_{\parallel})=\rho_{\pm,TE}(k_{\parallel}) \mp \sin(k_z h)/2$ \cite{kuang2020maximal}.\\

The same set of propagating plane-wave modes can also be used to expand the incident electric field as
\begin{equation} \label{Einc}
    {\bf{E^i}}= \frac{1}{k^{3/2}} \sum_{s,p} \int_{k_{\parallel}<k_i} e^i_{s,p}({\bf{k_{\parallel}}}) \, {\bf{v_{s,p}(k_{\parallel})}} \,  \frac{d {\bf{k_{\parallel}}}}{(2\pi)^2}.
\end{equation}
Without loss of generality, however, in the following we consider the incident field to be a single propagating plane wave, ${\bf{E^i}}= E_0 e^{i {\bf{k^i_{\parallel}}}\cdot {\bf{r_{\parallel}}} } e^{i {{k^i_{z}}}  z}\,{\bf{\hat{e}^i}}$, corresponding to choosing the expansion coefficients in Eq. (\ref{Einc}) as 
$e^i_{s,p}({\bf{k_{\parallel}}})= C(s,p)\, \sqrt{2 k^i_z k}\, E_0\, A \delta_{ {\bf{k_{\parallel}},k_{\parallel}^i}} \delta_{p,p^i}$. The values of C are as follows: $C(+,M)=-i,\, C(-,N)=i,\, C(+,N)=1,\, C(-,M)=-1 $. The last two values are multiplied by $-1$ if the propagation direction of the incident field is flipped from $z$ to $-z$ (backreflection).  The reflection vector ${\bf{F_s}}$ can also be expanded similarly.

Using these expansions, the terms in Eq. (\ref{eq:Uopt}) are evaluated as
\begin{equation} \label{eqUopt1}
    \beta= 2 A\, \cos(\theta^i) \frac{E_0^2}{k} \, \sum_s \frac{\rho_{s,p}(k^i_{\parallel})}{\textrm{Im} \zeta + \rho_{s,p}(k^i_{\parallel})},
\end{equation}
\begin{equation} \label{eqUopt2}
    \gamma= 2 A \,\cos(\theta^s) \frac{1}{k} \, \sum_s \frac{\rho_{s,p}(k^s_{\parallel})}{\textrm{Im} \zeta + \rho_{s,p}(k^s_{\parallel})},
\end{equation}
\begin{equation} \label{eqUopt3}
    \alpha= \frac{2 A E_0\, \cos(\theta^s)}{k} \, \delta_{ {\bf{k_{\parallel}^s},k_{\parallel}^i}}\, \delta_{p^s,p^i}     \sum_{s=\{+,-\}} s\, \frac{\rho_{s,p}(k^s_{\parallel})}{\textrm{Im} \zeta + \rho_{s,p}(k^s_{\parallel})},
\end{equation}
where the angle of incidence is $\cos(\theta^i)= k^i_{z}/k$, and the angle of reflection is $\cos(\theta^s)= k^s_{z}/k$. The first two terms characterize the incident and reflected waves independently, while the $\alpha$-term depends on the specific polarization and direction of both incident and scattered waves.  For a generic incident field, these terms consist of a summation over the $\bf{k_{\parallel}}$ spectrum of the incident wave.

\section{General reflection bound and validation}

The reliability of  Eq. (\ref{eq:Uopt}) as an upper bound on reflection from planar structures, and the validity of our assumptions, can be tested by comparing $U_{\textrm{opt}}$ to the standard Fresnel reflectance coefficient $R$ for a homogeneous film for both normal and oblique incidence cases, where $\theta^i =\theta^s$, and for both polarizations \cite{born2013principles}. For a physically consistent comparison between the far-field radiation intensity and the reflectance, a suitable normalization factor $U_0$ is introduced to ensure that $\tilde{U}_{\textrm{opt}}=U_{\textrm{opt}}/U_0$ does not exceed unity for passive systems {(we also note that, due to the finite extent of any physical structure, any measurement of reflectance actually measures the far-field radiation intensity rather than the ideal Fresnel reflectance from an infinite structure).} The reference case $U_0$ is the far-field radiation intensity, in the desired direction, from a perfect conductor, which ideally reflects all incident power. This can be calculated from Eq. (\ref{eq:U}) by evaluating the generated currents on a perfect conductor surface of area $A$, yielding $U_0=k^2A^2 E_0^2\, \cos^2\theta^i/(8\pi^2)$. Alternatively, the same result can be obtained by setting $\textrm{Im} \zeta \to 0$ in the bound expression above. For the general case with different incident and reflected directions, the normalization factor should be slightly modified as $U_0=k^2A^2 E_0^2\, \cos\theta^i\, \cos\theta^s/(8\pi^2)$.

Finally, the normalized upper bound $\tilde{U}_{\textrm{opt}}$ can be written as
\begin{equation} \label{eq:UoptR}
\begin{split}
\tilde{U}_{opt}&= \frac{1}{4} \Bigg(\Big| \delta_{ {\bf{k_{\parallel}^s},k_{\parallel}^i}}\, \delta_{p^s,p^i} \sum_{s=\{+,-\}} s\, \frac{\rho_{s,p}(k^i_{\parallel})}{\textrm{Im} \zeta + \rho_{s,p}(k^i_{\parallel})}\Big|+\\
&\sqrt{\sum_s \frac{\rho_{s,p}(k^i_{\parallel})}{\textrm{Im} \zeta + \rho_{s,p}(k^i_{\parallel})}\sum_s \frac{\rho_{s,p}(k^s_{\parallel})}{\textrm{Im} \zeta + \rho_{s,p}(k^s_{\parallel})}}\Bigg)^2.
\end{split}
\end{equation}

This expression, which represents the main result of the paper, provides a strict upper bound on reflection from a generic planar structure, for any polarization and direction of incidence and reflection. {The bound depends directly on the thickness of the structure, which determines the eigenvalues $\rho_{s,p}(k_{\parallel})$ , and on the loss factor $\textrm{Im}~ \zeta = \textrm{Im}~ \chi / |\chi|^2$. The bound converges to unity (perfect reflection) if the thickness diverges, $h \to \infty$, and/or in the lossless limit, $\textrm{Im}~ \zeta \to 0$. Indeed, if the material properties were unconstrained, it would be possible to create planar structures with ideal reflectance in any direction/polarization, as demonstrated by recent work on metasurfaces and meta-gratings \cite{asadchy2016perfect,ra2017metagratings,quan2019passive,alvarez2022fundamental}, by relying on perfect conductors or lossless plasmonic materials to create the desired (often resonant) response for any non-zero thickness or to engineer strong nonlocal effects (i.e., spatial dispersion) mediated by guided waves \cite{asadchy2016perfect,shastri2022nonlocal}. Instead, since real materials and metamaterials are always imperfect, i.e., they exhibit non-zero dissipation or scattering losses, it is very relevant to study the fundamental limits to reflection for real lossy materials.}

\begin{figure}[h!] 
\centering\includegraphics[width=9cm]{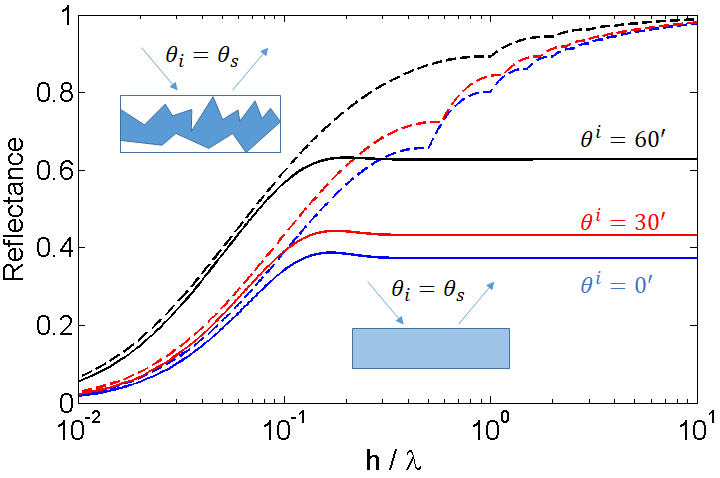}
\caption{ Reflectance $R$ of a homogeneous dielectric film (solid lines) vs. the upper bound $\tilde{U}_{\textrm{opt}}$ (dashed) on the specular reflectance of a generic structure  with the same thickness $h$ and the same loss factor $\textrm{Im}~ \zeta$. The incident field is a TE plane wave. The considered material for this example is Titanium Nitride with a refractive index $n=1.2+1.68i$ at $500$ nm \cite{rii}. The kinks in the curves for $\tilde{U}_{\textrm{opt}}$ are due to the absolute value that appears in the general solution, Eq. (\ref{eq:Uopt}), of the considered optimization problem. } \label{Fig2}
\end{figure}

{Figure  \ref{Fig2}}   shows a comparison between the Fresnel reflectance $R$ and the derived bound (\ref{eq:UoptR}) for a dielectric lossy film illuminated by a TE plane wave, as a function of the film thickness and for different incident angles. The polarization term $\delta_{p^s,p^i}$ equals unity since only reflection with the same polarization as the incident field is considered. For $h \ll \lambda$,  $\tilde{U}_{opt}$ is close to the Fresnel reflectance $R$, which indicates that a homogeneous film is the best design choice for the given dielectric lossy material to achieve maximum reflection in the deep-subwavelength region. This result is consistent with the fact that if the structure is thin compared to the wavelength, and the considered material is lossy, then patterning/structuring the thin film would not help, as the structure would not be able to induce any strong resonant response to maximize reflection.

The significance of the reflection bound becomes more apparent for larger thicknesses, as the bound becomes distinctly higher than $R$ (Figure \ref{Fig2}), and eventually converges to unity when the planar volume is several wavelengths thick. As the designable volume gets larger, it is more likely to find a (possibly resonant) current distribution that exceeds the homogeneous dielectric film reflection. Similar observations were made in Ref. \cite{kuang2020maximal}, but for the total scattering/absorption cross sections of thin films, and inverse-design methods were used to find structures that maximize absorption beyond the homogeneous film case. By repeating the comparison for different values of the refractive index and for both polarizations, it is found that, correctly, the bound always exceeds the Fresnel reflectance, i.e., $\tilde{U}_{opt} \geq R$, for any incidence angle, while the bound always increases with increasing thickness. Moreover, for vanishing thicknesses, $\tilde{U}_{opt}$ approaches $R$ and goes to zero for any non-zero loss factor $\textrm{Im}~ \zeta$.

\begin{figure}[h!] 
\centering\includegraphics[width=9cm]{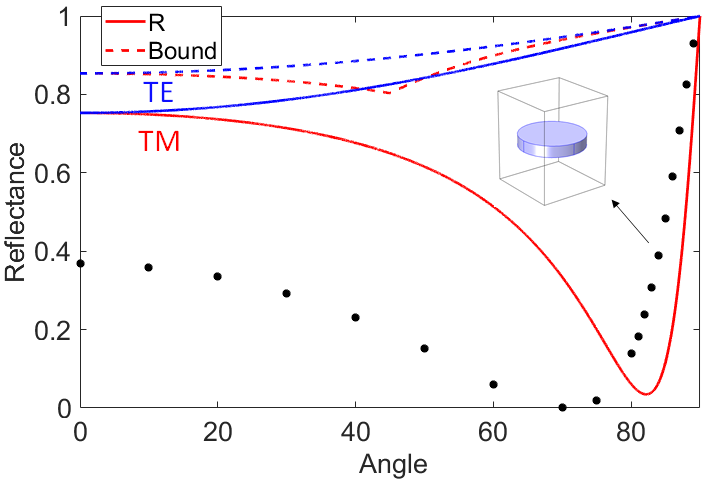}
\caption{ Reflectance $R$ of a homogeneous dielectric film (solid lines) vs. the reflection bound $\tilde{U}_{\textrm{opt}}$ (dashed), as a function of incidence angle, for both TE (blue) and TM (red) polarizations. The film thickness is $h/\lambda =0.04$  and the refractive index is $n=7+2i$. For the TM case, $R$ is suppressed around the Brewster angle, significantly deviating from the bound. The dots represent the specular reflection from a periodic square array of disks (the periodicity is $h/\lambda =0.2$) of the same height and material properties, which demonstrates five times the reflectance of the homogeneous film at the Brewster angle. COMSOL Multiphysics (v. 5.4) is used for the numerical evaluation of the reflected power for the array.} \label{Fig3}
\end{figure}

The case of TM polarization is more subtle and merits further discussion, as the bound significantly deviates from $R$ for large incidence angles. This is a consequence of the Brewster angle effect, according to which a TM wave is entirely transmitted through a homogeneous interface at the Brewster's angle $\theta_B= \tan^{-1}(n)$ \cite{orfanidis2002electromagnetic}. Intuitively, by disrupting the homogeneity of the thin-film through patterning/structuring, the Brewster angle condition is violated and therefore a significantly higher reflection, for the same incident angle, is certainly possible. In other words, the bound predicts that it is possible to find a configuration that yields a much higher reflection for the TM case near the Brewster angle, compared to a homogeneous film, even for deeply subwavelength thicknesses. This observation can be easily confirmed by simulating the reflection from a simple configuration different than a homogeneous film, such as a periodic array of disks of the same height and material properties, as shown in {Figure \ref{Fig3}}. 

\begin{figure}[h!] 
\centering\includegraphics[width=9cm]{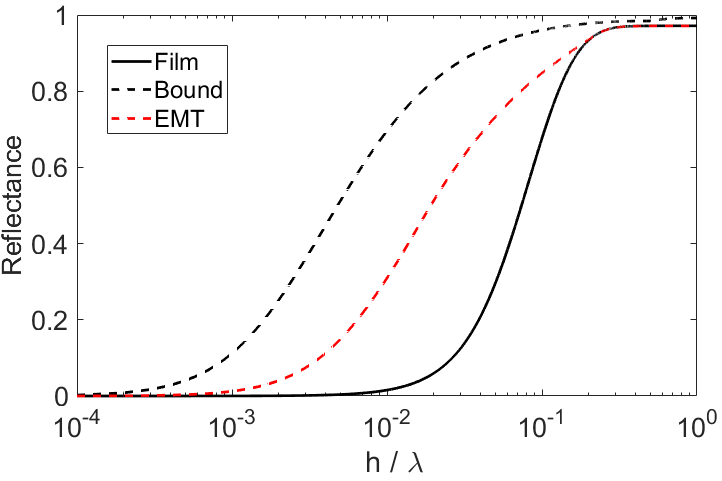}
\caption{ Comparison between the reflectance of a homogeneous plasmonic film of permittivity $\varepsilon_r=-3+0.1i$ (solid black line), the maximum possible reflectance of a homogenized-material film made of identical spherical inclusions with the same permittivity $\varepsilon_r$ (dashed red), and the upper reflection bound on a planar structure made of the same material (dashed black). All cases are for the same thickness $h$. The permittivity of the homogenized-material film is obtained using the Maxwell Garnett formula while varying the filling factor to determine the maximum possible reflectance for each value of thickness.} \label{Fig4}
\end{figure}

The upper reflection bound can also be validated by comparing it to the results of effective medium theories (EMTs), or homogenization theories, for an inhomogeneous distribution of matter. Homogeneous theories show that the collective behaviour of an ensemble of subwavelength, dipolar, polarizable particles, with subwavelength inter-particle distances, can be treated as the response of a homogeneous medium of well-defined effective macroscopic properties. {Specifically, the well-established Maxwell Garnett homogenization theory provides an analytical expression to determine the space of possible values of effective permittivity $\varepsilon_h$ that can be obtained from identical {spherical} inclusions of permittivity $\varepsilon_r$, as a function of the density of polarizable elements or filling factor \cite{maxwell1904colours,slovick2014generalized,scheller2010applications,belyaev2018electrodynamic}.} Maximum possible specular reflection from an homogenized film of thickness $h/\lambda$ can be then calculated by considering all the possible values of $\varepsilon_h$ predicted by the homogenization formula. As confirmed by {Figure \ref{Fig4}}, the derived upper reflection bound always exceeds the reflectance of an homogenized film. This is consistent with fact that the, while standard homogenization theories assume purely dipolar light-matter interactions, therefore constraining the resulting polarization current, our upper reflection bound does not assume any particular type of induced current and is therefore significantly more general.

\section{Results: Selected applications}

The fundamental bound on reflection derived in the previous section apply to any planar structure made of lossy materials, including homogeneous and homogenized slabs, multilayer thin films, gratings and meta-gratings, photonic crystal slabs, and metasurfaces. In this section, we discuss the use of these general bounds in three examples of application scenarios where designing efficient reflective structures is a crucial goal.

\subsection{Reflective mirrors for solar sails}

Solar and laser sails have been recently proposed for a variety of novel science missions
ranging from ultra-fast interstellar travel to imaging exoplanets via solar gravity lensing effects \cite{davoyan2021photonic}. This emerging technology requires minimizing the weight of the sail as a crucial requirement to increase the acceleration produced by the radiation pressure that propels the sail. Moreover, refractory materials are required since they can withstand high temperatures, which limits the range of available material properties. In this scenario, the design goal is to achieve a certain acceptable reflectance at optical wavelengths using realistic refractory materials and the minimum possible thickness (hence minimum weight). That is exactly the kind of problem that our derived reflection bound, given by Eq. (\ref{eq:UoptR}), can address. 

\begin{figure}[h!] 
\centering\includegraphics[width=9cm]{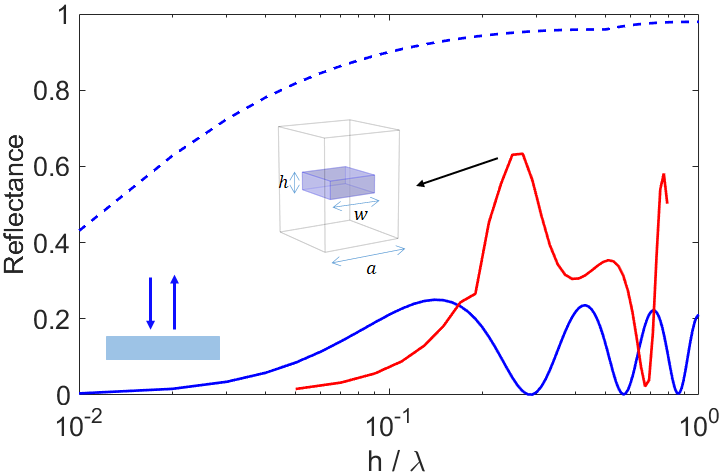}
\caption{ Normal-incidence reflectance $R$ of a homogeneous film (solid blue line), upper reflection bound (dashed blue), and reflectance of a periodic array of square prisms, optimized for a quarter-wavelength thickness. The considered material is $Al_2O_3$ with refractive index $n=1.75+0.02i$ at $500$ nm \cite{rii}.  } \label{Fig5}
\end{figure}

As an example, {Figure \ref{Fig5}}  shows that, for a lightweight low-loss refractory material like Aluminium Oxide $Al_2O_3$, a much higher reflectance is possible, beyond the homogeneous film case, even in the deeply subwavelength thickness regime. To demonstrate this point, the reflectance of a periodic square array of square prisms has been calculated and optimized using COMSOL 5.4, varying the width $w$ of the prisms and the lattice periodicity $a$ to achieve maximum reflection, for a thickness $h= \lambda/4$, at the center of the visible spectrum $\lambda = 500 $ nm. The optimized configuration ($w/\lambda=0.575, a/\lambda=0.95$) results in $64\%$ reflectance, which is almost three times the maximum possible reflectance from a homogeneous film with the same refractive index, as shown in Figure \ref{Fig5}. The bound suggests further enhancement is possible by considering more complicated structures designed using more advanced optimization techniques. 

Moreover, Table \ref{table:1} summarizes the minimum possible thickness for a reflective surface made of selected refractory materials to achieve at least $60\%$ and $90\%$ reflectance at $500$ nm. This analysis may be useful as a factor in the selection process of the optimal material for the design of solar and laser sails. We also note that another important question in the study of the fundamental limits of solar sails (powered by a broadband light source) is the issue of the maximum bandwidth over which a very high reflectance can be sustained. This will be the subject of future work.
  
 \begin{table}[h!]
\centering
\begin{tabular}{|c | c c |} 
 \hline
 Material & $R=60\%$ & $R=90\%$ \\ [0.5ex] 
 \hline\hline
 Al2O3 & 10 & 50  \\ 
 SiC & 15 & 85  \\
 TiN & 125 & 1050  \\
 W & 15 & 80  \\
 TiC & 40 & 348  \\ [1ex] 
 \hline
\end{tabular}
\caption{Minimum possible thickness (in nm) to realize a highly reflective planar mirror using typical refractory materials for solar/laser sails \cite{davoyan2021photonic}. The refractive indexes at $500$ nm are taken from \cite{rii}. }
\label{table:1}
\end{table}

\subsection{Reflection gratings }
\begin{figure}[h!] 
\centering\includegraphics[width=09cm]{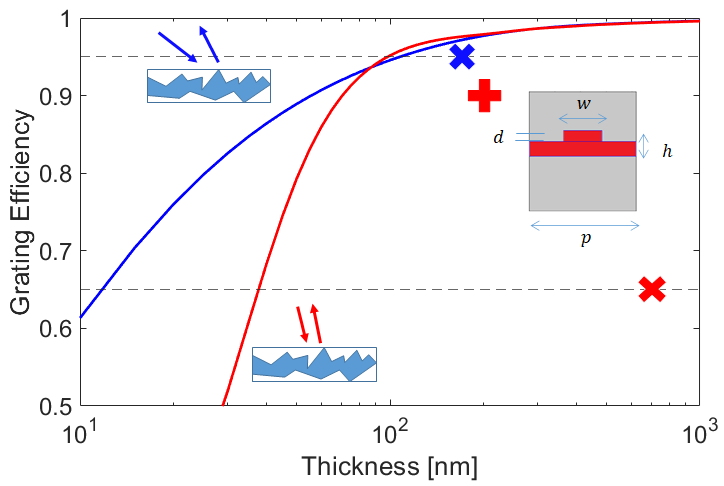}
\caption{Efficiency-thickness bounds on examples of reflective gratings. The blue 'X' marker indicates the efficiency of the binary gold grating in Ref. \cite{pgl}, which is close to the upper efficiency bound for its thickness (blue line). The red 'X' marker denotes the efficiency of the blazed aluminum grating in Ref. \cite{thor}, for which our upper bound (red line) suggests that better performance is possible. As an example, an optimized design for a binary aluminum grating is presented, marked by the red '$+$'. For this design, the achieved reflection is less than $10\%$ away from the upper bound.  The optimized  design parameters are  $w=300$ nm, and $d=85$ nm. The surrounding medium is vacuum, while the periodicity $p=833$ nm is chosen to reflect the incident light into the $-1$st order. The overall  thickness of the Al layer is $h=200$ nm, which is one-fourth the thickness of the design in \cite{thor}. } \label{Fig6}
\end{figure}

The derived reflection bound given by Eq. (\ref{eq:UoptR}) can also be used to determine the maximum possible efficiency of a reflection grating or meta-grating for a given diffraction order. For instance, the commercially available gold grating in Ref. \cite{pgl} is optimized to reflect $95\%$ of TM-polarized light at $800$ nm, incident at $\theta^i=54^{\circ}$, to the $-1$st reflection order at $\theta^s=-22^{\circ}$. As shown in {Figure \ref{Fig6}}  (blue `X'), this optimized grating with thickness $h=165$ nm is already close to the maximum possible efficiency (for that specific thickness) and to the minimum possible thickness (for that efficiency level) predicted by the bound for the considered material.

Another possible example is a commercially available aluminum blazed grating that is designed to reflect $65\%$ of an incident TE-polarized light in Littrow configuration
(back-reflection, $\theta^i=-\theta^s=10.36^{\circ}$) for UV wavelengths around $300$ nm \cite{thor}. The thickness of the Al coating is $800$ nm \cite{thorthick}. While the grating performance may be acceptable for commercial applications, Figure \ref{Fig6} (red `X') shows that there may still be significant room for designing a more efficient and compact grating for the given setup, material, and wavelength. For example, by varying the dimensions of a binary Al grating, we arrive at an optimized design (red `+' in Figure \ref{Fig6}) that reflects $90\%$ of the incident light with one-fourth the thickness of the original design.

\subsection{Polarization converters }

Polarization converters are an important class of electromagnetic/optical components that are widely used in several applications ranging from communications and imaging to molecular sensing \cite{cheng2014ultrabroadband, kang2017preserving, lonvcar2018reflective,chen2016review, feis2020helicity}. Assuming normally incident and reflected light, and specializing the bound for cross-polarized reflection by setting $\delta_{p^s,p^i}=0$, Eq. (\ref{eq:UoptR}) is reduced to a simplified form that directly limits the maximum relative amount of energy coupled to the orthogonal polarization in reflection:
\begin{equation}   \label{eq:UoptRPC}
\tilde{U}_{PC}= \frac{1}{4} \Big(\sum_s \frac{\rho_{s,p}(k^i_{\parallel})}{\textrm{Im}~ \zeta + \rho_{s,p}(k^i_{\parallel})}\Big)^2.
\end{equation}

{Interestingly, in the small-thickness limit, $h/\lambda \to 0$, this expression can be further simplified by using the small-argument approximation of trigonometric functions, yielding
\begin{equation}
	\tilde{U}_{PC} \approx \frac{1}{4} \Big(\frac{ \pi h/\lambda}{\textrm{Im}~ \zeta + \pi h/\lambda}\Big)^2,
\end{equation}	
for both TE and TM polarizations. Then, by also letting $\textrm{Im}~ \zeta \to 0$ (lossless limit), the bound converges to $\tilde{U}_{PC}\to 1/4$. This asymptotic result for the small-thickness lossless limit matches the theoretical limit on polarization cross-coupling derived earlier in the literature, from very different considerations, for any infinitesimally thin structure made of a passive lossless material \cite{monticone2013full,arbabi2017fundamental}. Intuitively, this can be explained from the fact that currents induced on a single ultra-thin layer (e.g., a metasurface made of thin, planar, lossless (nano)antennas) will necessarily radiate symmetrically toward both sides, resulting in an unavoidable 50\% reduction in reflection efficiency; in addition, the planar polarizable elements on the surface (e.g., obliquely oriented dipoles) can couple, at most, half of the incident field to the orthogonal polarization \cite{monticone2013full}, resulting in a maximum cross-polarized reflectance of $1/4$. This intuitive result is confirmed and generalized by our fundamental bound on reflection.}

\begin{figure}[h!] 
\centering\includegraphics[width=9cm]{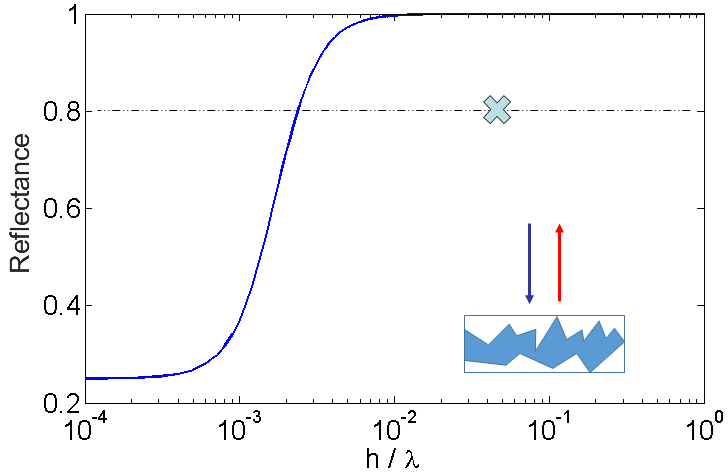}
\caption{Upper bound on the cross-polarized reflectance in the normal direction, as a function of thickness, for Cu at $10$ GHz. 
The `X' marker represents the reflectance obtained in Ref. \cite{lonvcar2018reflective} for a reflective polarization-converting metasurface. The comparison with the bound indicates that it may be possible to find structures with better polarization conversion performance and/or smaller thickness.} \label{Fig7}
\end{figure}

As an example, {Figure \ref{Fig7}}  shows the cross-polarized reflectance of a reflective microwave polarization converter from the literature \cite{lonvcar2018reflective} compared against the fundamental bound. Using a low-loss metal like Cu, it was possible to create a very thin and efficient polarization converter metasurface, with $R_{PC} \approx 0.8$ at $10$ GHz, given $h=1.27$ mm or $h/\lambda=0.04$ \cite{lonvcar2018reflective}. Nevertheless, our derived bound in Eq. (\ref{eq:UoptRPC}) suggests that a further minimization can be achieved. As illustrated in Figure \ref{Fig7}, the minimum possible thickness to achieve $R_{PC}= 0.8$, using the same material, is $h_{min}=0.003\lambda$, which is more than an order of magnitude lower than the proposed design in Ref. \cite{lonvcar2018reflective}. On the other hand, the bound indicates that there is no fundamental constraint that could prevent achieving a perfect polarization conversion since $\tilde{U}_{PC} \approx 1$ at $h/\lambda=0.04$. Rather, the main issue could be the computational resources required to find a better design and the feasibility to fabricate it. Finally, we note that the bound quickly reduces to approximately $1/4$ as the thickness is reduced, consistent with the discussion above.

\section{Conclusion}

In this paper, we have derived analytical upper bounds on the general problem of reflection from complex planar structures, irrespective of their specific design, and for any direction of incidence/reflection and polarization. The only assumptions are that the structure is passive, with a surface area much larger than its thickness, and is made of a single local, isotropic, and nonmagnetic material. The latter two assumptions may be relaxed, but the resulting bounds would no longer be expressible as simple closed-form formulas. We have validated our theoretical results by comparing the derived bounds against the standard Fresnel reflectance $R$ of homogeneous and homogenized, dielectric and plasmonic films, for both polarizations. For TM polarization, the bound predicts the possibility to achieve significantly higher reflection than $R$ near the Brewster angle, even for deeply subwavelength films, and we demonstrated this possibility with a numerical example. Interestingly, the  derived bound also replicates, confirms, and generalizes the previously derived limit on polarization conversion efficiency for very thin structures (25\% \cite{monticone2013full}). 

In the second part of the paper, we have applied the derived fundamental bound on reflection  to various results from the recent literature and the associated applications, focusing on ultra-thin reflective mirrors, reflection gratings, and polarization converters made of real, imperfect (i.e., lossy) materials. As a relevant example, we have identified the minimal possible thickness for several promising refractory materials for the design of ultra-light mirrors for solar/laser sails. Furthermore, while we have found that some designs are already optimized and result in performance very close to the bound, other designs can be further improved both in terms of compactness and reflectance. In this context, when using optimization techniques and inverse-design methods, the bound can serve as a guideline to minimize computational resources by identifying parameter regions where there is room for further enhancement. For future work, it would be interesting to explore the possibility to derive similar bounds for planar structures on substrates (e.g., metasurfaces on glass) or layered media, or more broadly in the presence of different scatterers, by using the Green's function expansion for a dipole in the considered environment. Our analysis may also be extended to include more complex material properties, such as magnetic and chiral materials. Moreover, more work is needed to extend these results to the problem of broadband maximization of reflection, establishing fundamental tradeoffs between bandwidth, thickness, and reflectance.

To conclude, given the importance of engineering and optimizing light reflection in many scenarios, we believe our results will prove useful to assess the performance of reflective components and provide relevant insight into how to improve their performance for various applications.

\medskip

\medskip
\textbf{Acknowledgements} \par 
The authors acknowledge support from the National Science Foundation (NSF) with Grant No. 1741694, and the Air Force Office of Scientific Research with Grant No. FA9550-22-1-0204 through Dr. Arje Nachman.

\medskip

%





\end{document}